\documentclass[aps,pra,notitlepage,nofootinbib,twocolumn]{revtex4-2}

\usepackage[english]{babel}
\usepackage[utf8x]{inputenc}
\usepackage{amsmath}
\usepackage{amsfonts}
\usepackage{amssymb}
\usepackage{bm}
\usepackage{physics}
\usepackage{xcolor}
\usepackage{algorithm}
\usepackage[noend]{algpseudocode}
\usepackage{hyperref}
\usepackage{graphicx}
\usepackage{tikz}
\usepackage{soul}
\hypersetup{colorlinks=true,citecolor=blue}

\begin{document}

\title{Bayesian Learning of Parameterised Quantum Circuits}
\author{Samuel Duffield}
\email{sam.duffield@cambridgequantum.com}
\affiliation{Quantinuum, Partnership House, Carlisle Place, London SW1P 1BX, United Kingdom}

\author{Marcello Benedetti}
\affiliation{Quantinuum, Partnership House, Carlisle Place, London SW1P 1BX, United Kingdom}

\author{Matthias Rosenkranz}
\email{matthias.rosenkranz@cambridgequantum.com}
\affiliation{Quantinuum, Partnership House, Carlisle Place, London SW1P 1BX, United Kingdom}

\date{June 15, 2022}

\begin{abstract}
Currently available quantum computers suffer from constraints including hardware noise and a limited number of qubits. As such, variational quantum algorithms that utilise a classical optimiser in order to train a parameterised quantum circuit have drawn significant attention for near-term practical applications of quantum technology. In this work, we take a probabilistic point of view and reformulate the classical optimisation as an approximation of a Bayesian posterior. The posterior is induced by combining the cost function to be minimised with a prior distribution over the parameters of the quantum circuit. We describe a dimension reduction strategy based on a maximum a posteriori point estimate with a Laplace prior. Experiments on the Quantinuum H1-2 computer show that the resulting circuits are faster to execute and less noisy than the circuits trained without the dimension reduction strategy. We subsequently describe a posterior sampling strategy based on stochastic gradient Langevin dynamics. Numerical simulations on three different problems show that the strategy is capable of generating samples from the full posterior and avoiding local optima.
\end{abstract}

\maketitle

\section{Introduction}

Variational quantum algorithms (VQAs)~\cite{McClean_2016,Benedetti_2019, cerezoVariationalQuantumAlgorithms2021,bhartiNoisyIntermediatescaleQuantum2022} are the leading paradigm for solving computational problems on current generation quantum computers.
A VQA solves the computational problem by turning it into an optimisation problem over the parameters of a quantum circuit. The quantum computer is used to execute the circuit, that is, to prepare a quantum state and perform measurements on it. The classical computer is used to estimate the cost function from measurement outcomes and to update the parameters accordingly. This process is repeated in the hope of finding the parameters yielding minimum cost, effectively encoding a solution to the computational problem.

\begin{figure*}[bht]
    \centering
    \includegraphics[width=6.0in]{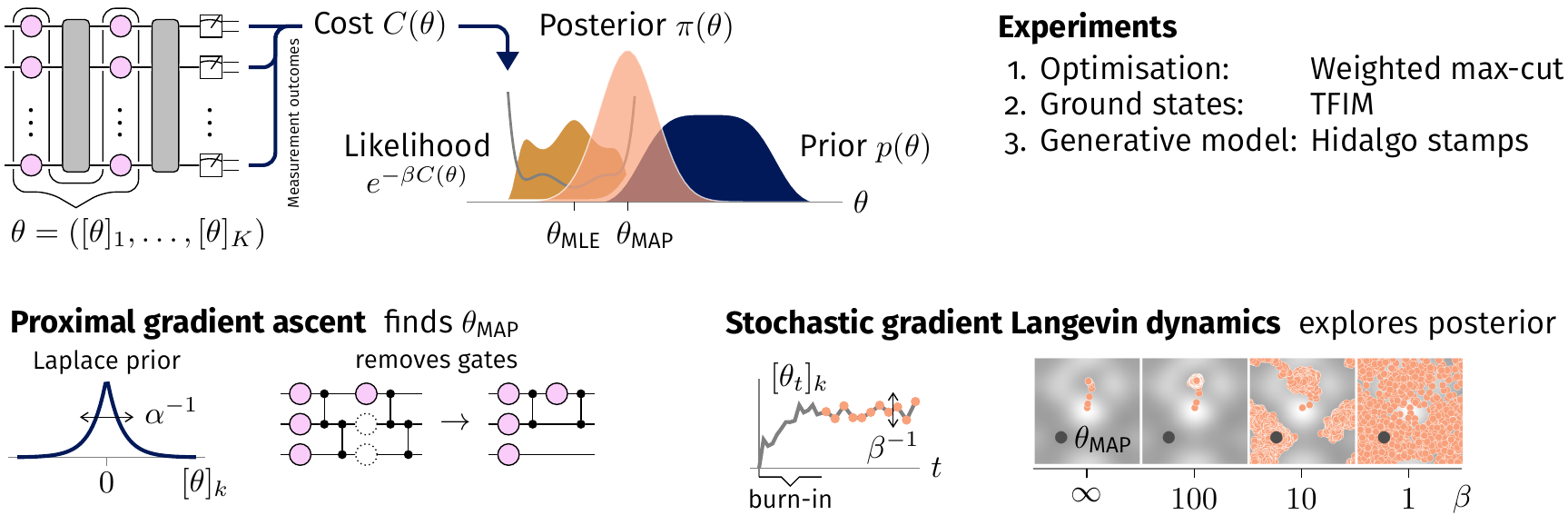}
    \caption{Bayesian perspective on learning parameterised quantum circuits. Circuit parameters $\theta$ define a likelihood term via a cost $C(\theta)$. A suitable choice of the cost function enables a variety of tasks, such as combinatorial optimisation, finding ground states of Hamiltonians, and generative modelling. The prior can be used to encode knowledge or desirable properties of the parameters. Typical goals are finding the mode of the posterior $\theta_\text{MAP}$ or exploring the full posterior. The former is achieved with proximal gradient ascent, which encourages gate count reduction. The latter is achieved with stochastic gradient Langevin dynamics, which can be useful for escaping local optima during training.}
    \label{fig:overview}
\end{figure*}

A number of VQAs have been proposed to attack specific problems in condensed matter physics, quantum chemistry, machine learning and combinatorial optimisation, with demonstrations on existing hardware~\cite{McClean_2016,Benedetti_2019, cerezoVariationalQuantumAlgorithms2021,bhartiNoisyIntermediatescaleQuantum2022}. One of the motivations is that the quantum circuit ansatz can be designed to comply with hardware constraints (e.g. qubit-to-qubit connectivity and coherence time) and to encode domain knowledge about the problem (e.g. symmetries and correlations). This is in contrast to fault-tolerant quantum algorithms which use a large number of error-corrected qubits and deep circuits to solve generic instances of a problem.  

Despite the successes of VQAs, it is well known that the optimisation of parameterised quantum circuits does not scale to large systems in general. The optimisation landscape is characterised by features such as barren plateaus~\cite{mccleanBarrenPlateausQuantum2018}, narrow gorges~\cite{cerezoCostFunctionDependent2021}, and exponentially many local minima~\cite{You_2021}, most of which have poor quality~\cite{anschuetzBarrenPlateausQuantum2022}. Real experiments are further complicated by the fact that hardware is noisy, and execution of quantum gates is slow in some architectures. Any improvement in parameter initialisation (e.g.~\cite{eggerWarmstartingQuantumOptimization2021,Nishant_2021,zhangGaussianInitializationsHelp2022}), ansatz design (e.g.~\cite{gardEfficientSymmetrypreservingState2020,congQuantumConvolutionalNeural2019,meyerExploitingSymmetryVariational2022}) and training (e.g.~\cite{skolikLayerwiseLearningQuantum2021,zhangTrainabilityDeepQuantum2021,Sack_2022}) could push the boundaries of VQA applications.

In the context of VQAs there exist a plethora of Bayesian methods which we briefly review here. Bayesian optimisation is a zeroth-order method (i.e. it does not use gradient information) which is popular among VQA practitioners~\cite{Otterbach2017,Zhu2019,Self2021}.
Bayesian optimisation can, however, be used within first-order methods to tune the stepsize \cite{Shiro2021} or initialisation \cite{Rad2022}. Reference~\cite{Wang_2021} uses Bayesian methods to infer the value of the cost function from a reduced number of measurements. Reference~\cite{Benedetti2021} proposes a VQA for inference of unobserved variables in Bayesian networks. Reference~\cite{Du2020} uses entanglement and ancillary qubits to implement a quantum prior distribution over the circuit parameters. Going beyond VQAs we also find a number of quantum algorithms for speeding up Bayesian inference~\cite{Low_2014, Harrow_2020} as well as novel Bayesian quantum causal models~\cite{Tucci_1995,leiferQuantumGraphicalModels2008,allenQuantumCommonCauses2017}. 

In this work, we formulate generic VQAs as a Bayesian inference problem over circuit parameters. We propose two algorithms that achieve different goals. One searches for the maximum a posteriori point estimate and automatically removes a given percentage of parameters (quantum gates) to reduce hardware noise and execution speed. The second approximately samples from the posterior distribution over the circuit parameters and reduces sensitivity to initialisation and local optima. Our methods make good use of the gradient of the cost function, which can be estimated from additional measurements on the quantum computer. An overview of the framework and methods is visualised in \figureautorefname~\ref{fig:overview}.
\par
This article is structured as follows. In Section~\ref{sec:bayes_persp} we introduce the Bayesian perspective for VQAs. In Section~\ref{sec:lap_prox} we describe the maximum a posteriori approach for a sparsity-inducing Laplace prior. In Section~\ref{sec:sgld} we describe stochastic gradient Langevin dynamics for posterior sampling. In Section~\ref{sec:sims} we numerically investigate the algorithms on instances of weighted max-cut, a transverse field Ising model, and a generative modelling problem. We also show the benefits of using the Laplace prior with an experiment on the Quantinuum H1-2 computer. In Section~\ref{sec:disc} we present our concluding remarks and discuss avenues for future research.

\section{A Bayesian Perspective}\label{sec:bayes_persp}

A parameterised quantum circuit (PQC) takes the form $U(\theta) = \prod_{k=1}^K  W_k U_k([\theta]_k)$, where $\{W_k\}_{k=1}^K$ is a set of fixed quantum gates, and $\{U_k([\theta]_k)\}_{k=1}^K$ is a set of parameterised gates. The circuit is applied to some initial quantum state. Let $C(\theta)$ be the cost for the problem at hand as a function of the circuit parameters $\theta = ([\theta]_1, \dots [\theta]_K)$. For example, the cost function could be the expectation value $C(\theta) = \Tr[O U(\theta) \ket{\psi_0}\bra{\psi_0} U(\theta)^\dagger]$ of an observable $O$ and initial state $\ket{\psi_0}$. In VQAs a classical optimiser is used in the hope of finding
\begin{equation}\label{mle}
    \theta_\text{MLE} = \text{argmin}_\theta C(\theta).
\end{equation}
Here we use the subscript to indicate that this quantity is akin to a \emph{maximum likelihood} estimator (MLE), as will become apparent. Assuming $C(\theta)$ is differentiable, a single iteration of vanilla (stochastic) gradient descent takes the form
\begin{equation}\label{GD}
    \theta_t = \theta_{t-1} - \epsilon_t \widehat{\nabla_\theta C}(\theta_{t-1}),
\end{equation}
where $\{ \epsilon_t \}_{t=1}^\infty$ is a schedule of stepsizes and $\widehat{\nabla_\theta C}(\theta)$ is an unbiased estimate of the gradient of the cost function $\nabla_\theta C(\theta)$~\cite{harrowLowdepthGradientMeasurements2021,swekeStochasticGradientDescent2020}.

Now let us consider a probabilistic formulation
\begin{equation}\label{posterior}
    \pi(\theta) = \frac{ p(\theta) \exp(- \beta C(\theta))}{\int p(\theta') \exp(- \beta C(\theta')) d\theta'},
\end{equation}
and generalised optimisation problem
\begin{equation}\label{map}
    \theta_\text{MAP} = \text{argmax}_\theta \pi(\theta).
\end{equation}
This probabilistic viewpoint instead treats the parameter vector $\theta$ as a random variable, and the denominator in Eq.~\eqref{posterior} ensures $\pi(\theta)$ is a valid probability distribution, i.e. $\int \pi(\theta)d\theta = 1$. In this light, we have the following \textit{Bayesian} interpretation:
\begin{itemize}
    \item $p(\theta)$ represents the \textit{prior distribution} and can be used to encode any pre-experimental knowledge or desirable properties for the parameters $\theta$;
    \item $\exp(- \beta C(\theta))$ is a generalised \textit{likelihood} term \cite{Bissiri2016, Pacchiardi2021} encouraging $C(\theta)$ to be small. The parameter $\beta$ controls the scaling of the cost function with respect to the prior. For statistical and machine learning tasks, the influence of the data enters through the likelihood term \mbox{(i.e. $C(\theta) = C(\theta; y)$ for a dataset $y$);} 
    \item $\pi(\theta)$ represents the \textit{posterior distribution} and describes high probability regions of the parameter space with uncertainty quantification.
\end{itemize}

The maximisation of $\pi(\theta)$ is equivalent to maximisation of $\log \pi (\theta)$ and thus we can take gradients in $\log$-space
\begin{equation}\label{log_grad_post}
    \nabla_\theta \log \pi(\theta) = \nabla_\theta \log p(\theta) - \beta \nabla_\theta C(
    \theta).
\end{equation}
We now observe that when the prior is set to the uniform distribution $p(\theta) \propto 1$, we get $\nabla_\theta \log p(\theta) = 0$ and $\theta_\text{MAP} = \theta_\text{MLE}$ from Eq.~\eqref{mle}. Subsequent application of a gradient ascent algorithm regains Eq.~\eqref{GD} (where the $\beta$ parameter is absorbed by the stepsize). Observe that the paradigm shift (going from minimising a cost function to maximising a posterior distribution) results in a change in terminology, gradient descent to gradient ascent - although in the case of a uniform prior the implementation is identical. This gradient ascent approach searches for a so-called \textit{maximum a posteriori} (MAP) estimator, as denoted in Eq.~\eqref{map} and described in generality in Algorithm~\ref{ga}  (note the stepsize has been rescaled such that $\beta \to \infty$ regains the maximum likelihood approach, Eq.~\eqref{GD}).

\begin{algorithm}[H]
\caption{Gradient Ascent}\label{ga}
\begin{algorithmic}
\For{$t = 1, \dots$}
\State $\theta_t = \theta_{t-1} + \epsilon_t \beta^{-1} \nabla_\theta \log p(\theta_{t-1}) - \epsilon_t \widehat{\nabla_\theta C}(\theta_{t-1})$
\EndFor
\end{algorithmic}
\end{algorithm}

A major success of the Bayesian paradigm is the ability to analyse uncertainty in the parameter $\theta$, that is to quantify the full posterior $\pi(\theta)$. Unfortunately, aside from trivial cases, the true posterior is intractable and we have to resort finding an approximation $q(\theta) \approx \pi(\theta)$. There then comes a trade-off between quality of approximation and computational cost. The cheapest approximation is that of the already discussed point estimate $q(\theta) = \delta(\theta \mid \theta_\text{MAP})$ where $\delta$ is the Dirac point measure, however this approach neglects all uncertainty in the parameter $\theta$. A more rigorous approach to approximating the posterior is to construct a Monte Carlo approximation $q(\theta) = \frac1T \sum_{t=1}^T \delta(\theta \mid \theta_t)$. The most popular methods for constructing this Monte Carlo approximation \cite{Brooks2011, Chopin2020} do so in a way that is asymptotically unbiased for the posterior, i.e. $q(\theta) \overset{T \to \infty}{\longrightarrow} \pi(\theta)$. Naturally, taking $T\to \infty$ is not feasible in practice and instead finite sample sizes are used, a (controllable) bias is therefore induced. A final approach to approximating the posterior is that of variational inference \cite{Blei2017}, where a parameterised \textit{variational family} of distributions $Q = \{q_\omega(\theta) : \omega \in \Omega\}$ is defined and then the optimal parameters $\omega^*$ are sought in order to minimise some tractable measure of the discrepancy between $q_\omega(\theta)$ and $\pi(\theta)$ (most commonly the KL divergence). Variational inference is typically computationally cheaper than the Monte Carlo approach, although induces a bias in the likely case $\pi(\theta) \notin Q$ and this bias can be difficult to assess or control.
\par
The Bayesian paradigm also provides a natural approach to characterising predictions, that is through expectations with respect to the posterior distribution
$
    \mathbb{E}_{\pi(\theta)}[f(\theta)] \approx \mathbb{E}_{q(\theta)}[f(\theta)],
$
where $f(\theta)$ is some predictive function. These predictions are trivial to implement in the case of the point estimate $\mathbb{E}_{q(\theta)}[f(\theta)] = f(\theta_\text{MAP})$ and Monte Carlo $\mathbb{E}_{q(\theta)}[f(\theta)] = \frac1T \sum_{t=1}^T f(\theta_t)$ approximations. In the case of variational inference, the variational family is often chosen such that $\mathbb{E}_{q_\omega(\theta)}[f(\theta)]$ is analytically tractable for the predictive functions of interest.

\section{Laplace Prior and Proximal Gradient Ascent}\label{sec:lap_prox}

A simple and relevant choice of prior is that of the Laplace distribution (independent across parameters)
\begin{equation*}
    p(\theta) \propto \exp \left( -
    \alpha \sum_{k=1}^K | [\theta]_k |
    \right),
\end{equation*}
for $\alpha \in [0, \infty)$. This choice of prior is also known as LASSO or $\ell_1$ regularisation \cite{Tibshirani1996}. For $\alpha$ large enough it is known that the resulting MAP estimate enforces $[\theta]_k = 0$ for the least influential parameters \cite{Tibshirani1996, Parikh2014}. In typical PQCs, the parameter $[\theta]_k$ represents an angle of a rotational gate and therefore setting $[\theta]_k = 0$ is equivalent to removing the corresponding gate. Removal of parameterised gates may lead to further gate reductions in a compilation step, e.g. if the removed gate was sandwiched between 2-qubit gates that now evaluate to the identity (see inset ``Proximal gradient ascent'' in Fig.~\ref{fig:overview}).

The non-differentiability of the Laplace prior also means we cannot apply standard gradient techniques. Specifically, Algorithm~\ref{ga} fails to enforce the parameters to be exactly zero. Fortunately we can utilise well-studied proximal gradient methods \cite{Parikh2014}. A single step of proximal gradient ascent takes the form
\begin{align*}
    \theta_t &= \text{prox}^\varphi_{\alpha \epsilon_t} \left(\theta_{t-1} - \epsilon_t \nabla_\theta C(\theta_{t-1})\right),
\end{align*}
where the proximal operator is defined as $[\text{prox}^\varphi_{\upsilon}(x)]_k = \text{argmin}_{y} \left\{ \varphi(y) + \frac{1}{2\upsilon} \| [x]_k - y\|^2 \right\}$. In general the proximal operator is intractable, however the special case of the Laplace prior $\varphi(y) = \ell_1(y) := |y|$ can be solved analytically giving the \textit{soft-thresholding} function \cite{Combettes2005}
\begin{equation}\label{prox}
    [\text{prox}^{\ell_1}_{\upsilon}(\theta)]_k = 
    \begin{cases}
    [\theta]_k - \upsilon, \quad & [\theta]_k > \upsilon, \\
    0,  & -\upsilon \leq [\theta]_k \leq \upsilon, \\
    [\theta]_k + \upsilon, &   [\theta]_k < - \upsilon. \\
    \end{cases}
\end{equation}
Proximal gradient ascent has the benefit of explicitly setting parameters $[\theta]_k = 0$ when they are sufficiently small, it is also known to converge to a (local) MAP estimate of $\pi(\theta) \propto \exp(-\alpha \sum_{k=1}^K \varphi([\theta]_k) - \beta C(\theta))$ and at a faster rate than vanilla gradient ascent \cite{Parikh2014} (which is regained, for the MLE, by setting $\alpha = 0$). We note that the influence of the $\beta$ parameter is absorbed by rescaling the regularisation parameter $\alpha$ and stepsize $\epsilon_t$ and can therefore be omitted, as described in Algorithm~\ref{pga}, where we also allow the parameter $\alpha=\alpha_t$ to adapt over iterations.

\begin{algorithm}[H]
\caption{Proximal Gradient Ascent}\label{pga}
\begin{algorithmic}
\For{$t = 1, \dots$}
\State $\theta_t = \text{prox}^{\varphi}_{\alpha_t \epsilon_t}(\theta_{t-1} - \epsilon_t \widehat{\nabla_\theta C}(\theta_{t-1}))$
\State
\hfill
for a Laplace prior, $\text{prox}^{\varphi}_{\upsilon}(\theta) = \text{prox}^{\ell_1}_{\upsilon}(\theta)$ in Eq.~\eqref{prox}
\EndFor
\end{algorithmic}
\end{algorithm}

There are several potential benefits of the Laplace prior and the subsequent reduction in the number of gates. Specifically, the resulting VQA benefits from
\begin{itemize}
    \item Reduced hardware noise;
    \item Circuits that are faster to sample.
\end{itemize}
The second point is also applicable to classical neural networks, where \textit{weight pruning}~\cite{Williams1995} is used to reduce time and memory costs at test time.
\par
There may also be trainability benefits in some cases (perhaps mitigating the barren-plateau phenomena \cite{mccleanBarrenPlateausQuantum2018} or simpler optimisation via dimension reduction) or better generalisation (by avoiding overfitting) for machine learning tasks, although this will be very case dependent as is investigated with mixed results in \cite{Qian2021} (where they use \textit{weight decay} which is equivalent to a Laplace prior without the explicit removal of gates achieved by proximal gradient ascent).
\par
The downside is that by maximising $\pi(\theta)$ rather than minimising $C(\theta)$ directly, we have $C(\theta_\text{MAP}) > C(\theta_\text{MLE})$ when the prior is not uniform. We theoretically find a worse solution, although if the circuit is deep and $\alpha$ is small, this difference in cost may be negligible.
\par
Additionally, the proximal operator is not tractable in general (the Laplace distribution is a particular instance where it is \cite{Polson2015}). This makes inference difficult for alternative priors such as the spike-and-slab \cite{Ishwaran2005} or horseshoe \cite{Carvalho2009} which theoretically achieve dimensionality reduction with a less significant shift in the global optima.

In practice, it is difficult to set the regularisation parameter $\alpha$. It is more intuitive to set a fixed percentage of the parameters $\theta = ([\theta]_1, \dots [\theta]_K)$ to be zero and have the algorithm adapt $\alpha$ (or rather $\alpha_t$) accordingly. We can achieve this at each iteration of proximal gradient ascent by choosing
\begin{equation*}
    \alpha_t \quad \text{such that} \quad \sum_{k=1}^K \mathbb{I}\left(\left|\left[\theta_{t-\frac12}\right]_k \right| < \alpha_t \epsilon_t \right) = K_0,
\end{equation*}
where $\theta_{t-\frac12} = \theta_{t-1} - \epsilon_t \widehat{\nabla_\theta C}(\theta_{t-1})$ and $K_0 \in \{1,\dots, K\}$ is the desired number of parameters to be set to 0 and therefore removed from the circuit. Numerically, this can be done efficiently by setting $\alpha_t$ to be the $\frac{K_0}{K}$th quantile of $\{|[\theta_{t-\frac12}]_k|/\epsilon_t \}_{k=1}^K$ at each iteration.

\section{Stochastic Gradient Langevin Dynamics}\label{sec:sgld}

Reducing the entire posterior $\pi(\theta)$ to a single point estimate $\theta_\text{MAP}$ will be a poor description of the true behaviour of the parameter $\theta$ unless the posterior is very concentrated. Furthermore, the gradient ascent algorithm may only succeed in finding a local maximum. A more desirable inference procedure characterises the entire distribution $\pi(\theta)$.
\par
The most popular classical approaches to posterior quantification build a Monte Carlo approximation normally either through Metropolis-Hastings based Markov chain Monte Carlo~\cite{Brooks2011} or importance sampling~\cite{Chopin2020}. Unfortunately both of these techniques require access to pointwise evaluations of $\pi(\theta)$ or at least, an unbiased estimate \cite{Andrieu2009}. In our setting, we only have unbiased estimates of $\log \pi(\theta)$ and $\nabla_\theta \log \pi(\theta)$. Fortunately, we can adopt the \textit{stochastic gradient Langevin dynamics} (SGLD) method of \cite{Welling2011} to generate an asymptotically unbiased Monte Carlo approximation whilst staying entirely in $\log$-space.
\par
\textit{Langevin dynamics} are described by the following stochastic differential equation
\begin{equation*}
    d\theta_t = \nabla_\theta \log \pi(\theta_t) dt + \sqrt{2}W_t,
\end{equation*}
where $W_t$ is a standard Brownian motion. Langevin dynamics are known to admit $\pi(\theta)$ as a stationary distribution \cite{Ma2015}. That is, if we take a sample $\theta_{t} \sim \pi(\theta)$ and evolve it exactly according to Langevin dynamics (for any time period $\epsilon$) then the marginal distribution of $\theta_{t+\epsilon}$ will also be $\pi(\theta)$. Therefore, simulating Langevin dynamics exactly and collecting samples along the way will provide a Monte Carlo approximation of the distribution $\pi(\theta)$.
\par
For non-trivial distributions Langevin dynamics cannot be simulated exactly. Instead, an Euler-Maruyama discretisation is commonly applied
\begin{equation}\label{langevin_em}
    \theta_t = \theta_{t-1} + \epsilon_t \nabla_\theta \log \pi(\theta_{t-1}) + \sqrt{2 \epsilon_t}\xi_t, \qquad \xi_t \sim \mathcal{N}(\xi \mid 0, \mathbb{I}).
\end{equation}
This discretisation will introduce a bias for practical stepsizes $\epsilon_t > 0$. However, as argued in \cite{Welling2011} (and proved in \cite{Teh2016}), as long as the stepsize schedule is chosen to decay to zero $\sum_{t=1}^\infty \epsilon_t^2 < \infty$ but not too fast $\sum_{t=1}^\infty \epsilon_t = \infty$, then the samples will be asymptotically correct for $\pi(\theta)$. It was also noted that this is the case if $\nabla_\theta \log \pi(\theta)$ is replaced with an unbiased estimate, thus obtaining \textit{stochastic gradient} Langevin dynamics. As a result, we can use our unbiased gradient estimate (or even a mini-batched version if applicable) within an SGLD algorithm to obtain a Monte Carlo approximation to the posterior $\pi(\theta)$. The algorithm, described in Algorithm~\ref{sgld}, represents a modification of gradient ascent with the correct amount of noise added to ensure exploration.

\begin{algorithm}[H]
\caption{Stochastic Gradient Langevin Dynamics}\label{sgld}
\begin{algorithmic}
\For{$t = 1, \dots$}
\State $\xi_t \sim \mathcal{N}(\xi \mid 0, \mathbb{I})$
\State $\theta_t = \theta_{t-1} + \epsilon_t \beta^{-1} \nabla_\theta \log p(\theta_{t-1}) - \epsilon_t \widehat{\nabla_\theta C}(\theta_{t-1})$
\State \hfill $ + \sqrt{2\beta^{-1}\epsilon_t}\xi_t$
\EndFor
\end{algorithmic}
\end{algorithm}

There are multiple potential benefits to adding noise to gradient steps in this principled manner. Specifically, the resulting VQA benefits from
\begin{itemize}
    \item Transitions between local optima and saddle points, for a suitably chosen $\beta$;
    \item Reduced sensitivity to initialisation;
    \item By replacing point estimate predictions with ergodic averages over the full trajectory (i.e. $\frac1T \sum_{t=1}^T f(\theta_t)$ for a prediction function $f(\theta)$), we quantify uncertainty in the parameter $\theta$. This is both conceptually advantageous and can provide enhanced generalisation and a reduced tendency to overfit for machine learning problems \cite{Izmailov2018}.
\end{itemize}
We also note that, unlike the proximal gradient approach, SGLD is extremely flexible to prior specification.

On the flip side, SGLD brings an additional parameter to tune, $\beta$. Setting $\beta$ too low will result in noisy trajectories that do not successfully find low cost regions of the parameter space. Conversely $\beta \to \infty$ regains vanilla gradient ascent with uniform prior, missing out on the benefits described above. Additionally, quantifying the posterior (via ergodic averages) is a significantly more challenging computational task than a simple point estimate and therefore more iterations may be required.

\section{Experiments}\label{sec:sims}

We now investigate the benefits, drawbacks and parameter sensitivities of the two aforementioned generalisations of gradient ascent in a selection of VQA experiments. Firstly, we examine an 11 qubit weighted max-cut  problem; secondly, we study the problem of sampling the ground state of an 11 qubit transverse-field Ising model; before finally exploring the statistical task of using an 8 qubit PQC as a generative model (a so-called quantum \textit{Born machine}~\cite{Cheng_2018}) for a real life, integer data set.
\par
One of the simplest circuit parameterisations is via single-qubit gates $\smash{U_k([\theta]_k) = e^{-i [\theta]_k V_k / 2}}$. These are rotations through angles $[\theta]_k \in [0, 2\pi]$, generated by Hermitian operators $V_k$ with eigenvalues $\pm1$. When the cost function $C(\theta)$ can be expressed as the quantum expectation of an Hermitian observable, the partial derivatives can be evaluated from parameter-shifted circuits (e.g. \cite{Li2017})
\begin{align*}
    [\nabla_\theta C(\theta)]_k = \frac12 \left( C(\theta + \tfrac{\pi}{2} e_k) - C(\theta - \tfrac{\pi}{2} e_k) \right) ,
\end{align*}
where $e_k$ is the unit vector in the $k$th direction.
\par
On a quantum device both $C(\theta)$ and each $[\nabla_\theta C(\theta)]_k$ can be approximated (unbiasedly) by generating $n_\text{shots}$ samples from the PQC for each expectation, however for our numerical experiments we make use of JAX \cite{jax} for exact cost and gradient evaluations \mbox{(i.e. $n_\text{shots} = \infty$).}
\par
All simulations are repeated 20 times with new initial parameters sampled from a small perturbation around zero, $[\theta_0]_k \sim \mathcal{U}([\theta]_k \mid -r, r)$ independently for $k=1,\dots, K$, with $r = 10^{-3}$. For each experiment we use a decaying stepsize schedule $\epsilon_t = a (t + b)^{-\frac13}$ in line with \cite{Teh2016}, and set $a=15$, $b=10$.

\begin{figure}
    \centering
    \begin{minipage}{\linewidth}
        \centering
        \scalebox{0.8}{
        \input{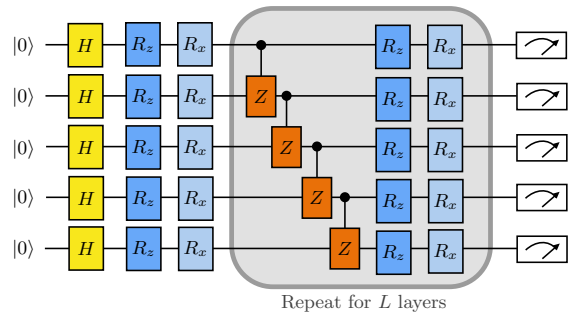}
        }
        \caption{PQC used for experiments illustrated for $N=5$ qubits. Each $R_x$ and $R_z$ gate is accompanied by a parameter $[\theta]_k$, $H$ is the Hadamard gate and all entangling gates are controlled $Z$ gates ($CZ$).}
        \label{fig:circ}
    \end{minipage}
\end{figure}

\begin{figure*}[bht]
    \centering
    \includegraphics[width=\linewidth]{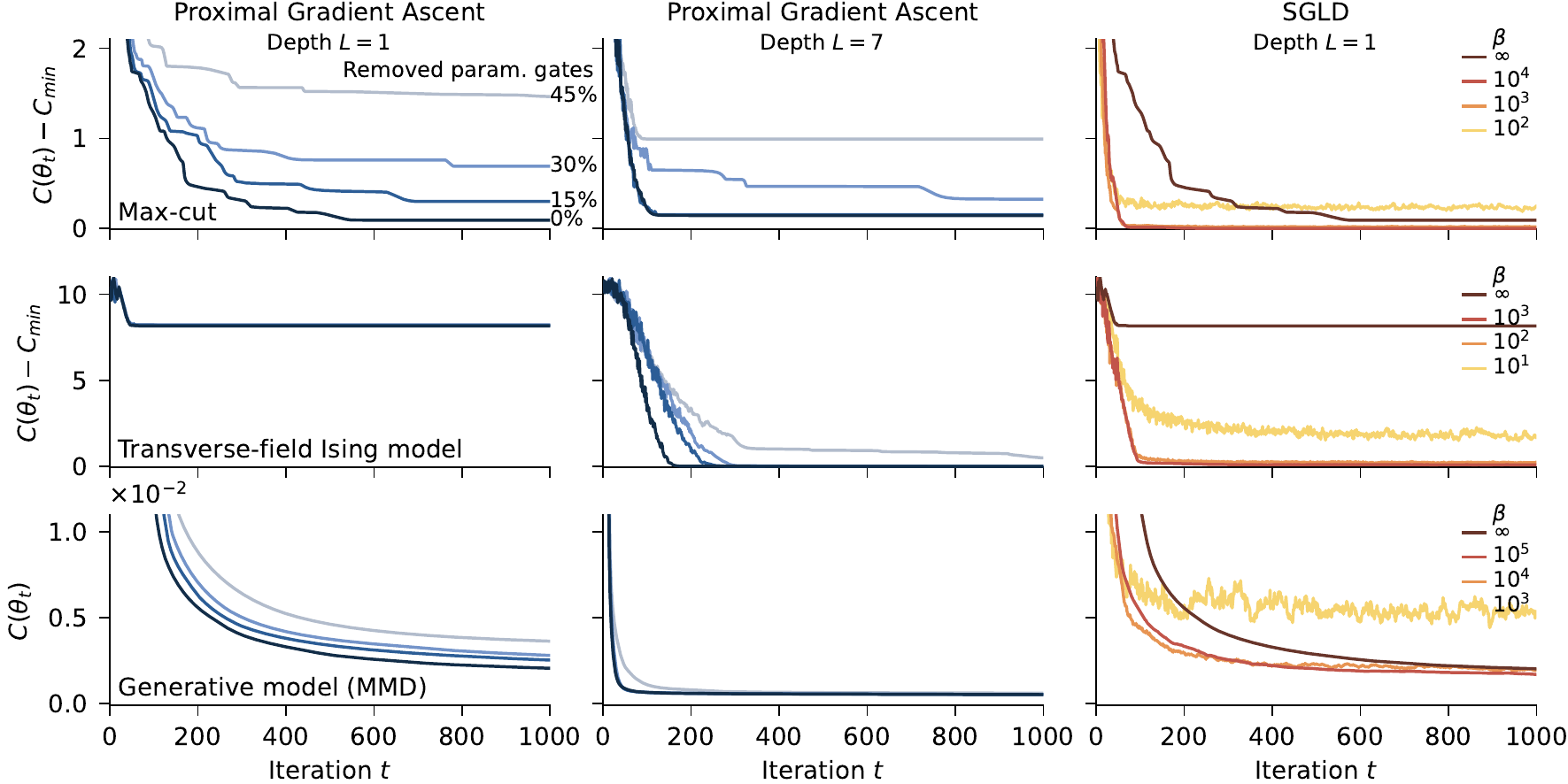}
    \caption{Training convergence of presented algorithms on weighted max-cut  (top row, 11 qubits), transverse-field Ising model (middle, 11 qubits) and generative modelling (bottom, 8 qubits) experiments. Proximal gradient ascent with Laplace prior, adaptive $\alpha_t$ with different percentages of parameterised gates automatically removed, and circuit depth $L=1$ (left column), $L=7$ (middle column), respectively. SGLD with depth $L=1$, uniform prior and varying noise levels $\beta$ (right column). All experiments are repeated across 20 random seeds with median displayed. Costs are shifted by their true minimum $C_\text{min}$. Circuit parameters are initialised with a small perturbation about zero and exact gradients are used ($n_\text{shots} = \infty$).}
    \label{fig:conv_plots}
\end{figure*}

\subsection{Weighted Max-cut}\label{maxcut}

A well-known NP-complete optimisation problem is weighted max-cut. This is the task of taking a graph of nodes and weighted edges then solving for the optimal binary labelling of the nodes. Say we label each node as either `0' or `1' with the first node fixed to be `0', then the optimality of the labelling is defined as maximising the sum of weights on edges between nodes with \textit{differing} labels. The cost function which we look to minimise is therefore defined as
\begin{equation*}
    C(\theta) = \mathbb{E}_{p(z \mid \theta)}[-S(z)],
\end{equation*}
where $z\in \{0, 1\}^N$ is a bit string labelling the $N + 1$ nodes (the zeroth node is labelled 0 by default), and $S(z)$ is the sum of weights between nodes with differing labels after labelling the graph according to the bit string $z$ (the $j$th element of $z$ indicates the label for the $j$th node).

We map bit strings to measurement operators in the computational basis of $N$ qubits as $z \rightarrow \dyad{z}{z}$. Then, the probability distribution is given by the Born rule $p(z\mid \theta) = | \bra{z} U(\theta) \ket{0}^{\otimes N} |^2$. This arises from the inherent randomness of the pure quantum state and depends on the parameter values $\theta$. For this experiment, we use the PQC in \figureautorefname~\ref{fig:circ} with $N=11$. The experiments are repeated across the 20 random seeds where for each seed a 3-regular graph (each node has three connected edges) is randomly generated along with associated weights each sampled uniformly in [0,1].

The top row of \figureautorefname~\ref{fig:conv_plots}, displays training with a Laplace prior via proximal gradient ascent on shallow and deep circuits as well as training with a uniform prior via SGLD on a shallow circuit.
\par
We vary the regularisation strength of the Laplace prior by changing the number of parameters set to 0 (and therefore removing gates from the circuit) at each iteration within the adaptive proximal gradient ascent, Section~\ref{sec:lap_prox}. For the shallower circuit ($L=1$ layers) we see that the Laplace prior is having a strong impact, forcing the trained cost to be higher. This is somewhat remedied by using a deeper circuit ($L=7$ layers).

In the SGLD plot we use the shallower circuit ($L=1$ layers). We observe that vanilla gradient ascent ($\beta=\infty$) is getting caught in local optima and that this is avoided by adding a suitable amount of noise in SGLD ($\beta=10^3$ and $\beta=10^4$), however adding too much noise ($\beta = 10^2$) prevents convergence to low cost regions. 

\begin{figure}
    \centering
    \includegraphics[width=\linewidth]{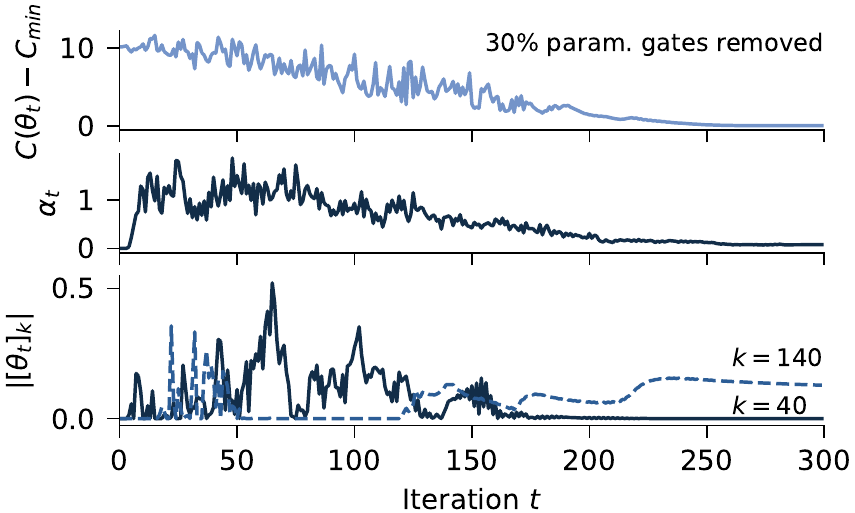}
    \caption{Proximal gradient ascent for finding the ground state of an 11-qubit transverse-field Ising model with a PQC of depth $L=7$. Training cost shifted by true ground state energy (top), adaptation of $\alpha_t$ parameter for target of 30\% parameterised gates removed (middle) and paths for two of the parameters $[\theta]_k$ (bottom). The two parameters (bottom) are chosen to illustrate that parameters are able to both enter and escape the zero threshold during training.
    }
    \label{fig:ad_alpha}
\end{figure}

\subsection{Transverse-field Ising model}\label{qh}
In the second experiment we are tasked with finding the ground state of a transverse-field Ising model (TFIM) with nearest neighbour interactions
\begin{equation}\label{eq:TFIM}
    H = - \sum_{i=1}^{N-1} Z_i Z_{i+1} - g \sum_{i=1}^N  X_i.
\end{equation}
This is a model of quantum magnetism, where $g$ corresponds to the applied transverse field in units of the Ising coupling strength~\cite{sachdevQuantumPhaseTransitions2011}. For the experiments, we choose $N=11$ and generate random TFIM instances by sampling $g \sim \mathcal{N}(\cdot \mid 0, \frac14)$. In the classical limit ($g=0$) the ground state is either all spins up or down, and the system remains in a ferromagnetic phase for $|g|<1$. It undergoes a phase transition at a critical point $|g|=1$ and remains in a disordered phase for $|g|>1$~\cite{sachdevQuantumPhaseTransitions2011}. The natural cost function for finding the ground state of Eq.~\eqref{eq:TFIM} with a PQC is
\begin{equation*}
    C(\theta) = \bra{0}^{\otimes N} U(\theta)^\dag H U(\theta) \ket{0}^{\otimes N}.
\end{equation*}
\par
We run the same setup as in the weighted max-cut experiment and display the training results in the middle row of Figure~\ref{fig:conv_plots}. We again observe severe local optima behaviour in the shallow circuit but this time the deep circuit successfully trains and finds the ground state even with 30\% of the gates removed from the circuit. We also notice that the addition of noise in SGLD is very effective at escaping the local optima, although it does not find the exact ground energy, most likely due to the reduced expressivity of the shallow circuit versus the deep circuit.
\par
For this experiment, we also visualise the adaptation of the Laplace prior regularisation parameter as described in Section~\ref{sec:lap_prox}. In Figure~\ref{fig:ad_alpha}, for a specific instance of the quantum Hamiltonian we observe that the regularisation parameter $\alpha_t$ quickly becomes large and then decreases as the algorithm converges to low cost regions of the optimisation landscape, where at each iteration 30\% of the parameters are set to zero. We additionally see that the set of zero parameters changes during training---as some parameters are dragged within the zero threshold this allows others to leave.

\subsection{Generative modelling}\label{gen_mod}
\begin{figure}
    \centering\includegraphics[width=\linewidth]{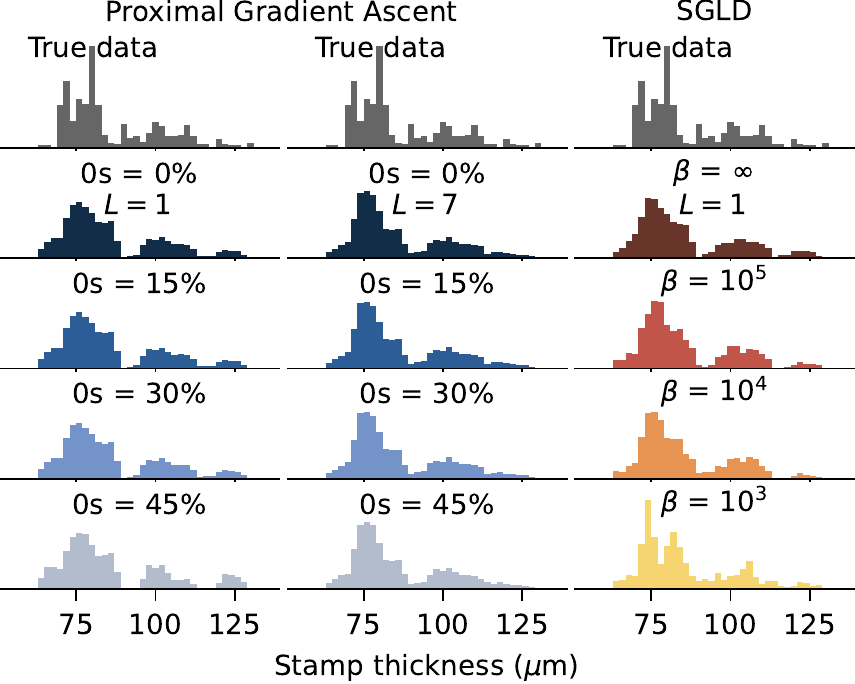}
    \caption{Distribution of true (top row) and simulated stamp data (bottom rows) sampled from MMD-trained quantum Born machines. Proximal gradient ascent with Laplace prior, adaptive $\alpha_t$ with different percentages of parameterised gates automatically removed, and circuit depth $L=1$ (left column), $L=7$ (middle). SGLD with varying $\beta$ and circuit depth $L=1$ (right column).}
    \label{fig:stamps_sim_data_plots}
\end{figure}

In our final experiment, we use the PQC in Figure~\ref{fig:circ} as a generative model (or Born machine) for the 1872 Hidalgo stamp dataset \cite{Izenman1988}. This dataset, $y$, represents measurements of the thickness of 485 stamps, on the $\mu$m scale these measurements are integers ranging from 60$\mu$m to 131$\mu$m and thus an 8-qubit generative model is sufficient to model in the binary expansion. This stamp data is displayed in the histograms on the top row of Figure~\ref{fig:stamps_sim_data_plots}.
\par
The goal of this generative modelling experiment is to drive the samples from the Born machine to be as close as possible to the true data set. The \textit{ideal} posterior is $p(\theta \mid y) \propto p(\theta)\prod_{j=1}^{485} p([y]_j \mid \theta)$ where $[y]_j$ is a single datum of the dataset $y = \{[y]_j\}_{j=1}^{485}$ and $p([y]_j \mid \theta)$ is the probability of the Born machine generating said datum for given parameters $\theta$. Pointwise evaluations of this likelihood $p([y]_j \mid \theta)$ are inherently intractable by the nature of quantum computation, and thus we cannot use it within our cost function. Instead, we utilise a generalised Bayesian inference framework \cite{Knoblauch2019}, replacing the true loglikelihood with a two-sample test or scoring rule \cite{Pacchiardi2021} that provides a measure of distance between a sample generated by the Born machine and the true data. In particular, we use the \textit{maximum mean discrepancy} (MMD) \cite{Gretton2012,Liu2018}
\begin{multline*}
    C(\theta) = \mathbb{E}_{p(z \mid \theta)p(z' \mid \theta)}[k(z,z')] \\
    - 2 \mathbb{E}_{p(z \mid \theta)\nu(z')}[k(z,z')]
    + \mathbb{E}_{\nu(z)\nu(z')}[k(z,z')],
\end{multline*}
where $\nu(z) = \frac{1}{485} \sum_{j=1}^{485} \delta(z \mid y_j) $ is the empirical distribution representing the dataset $y$. Here $k(z,z')$ is a kernel measuring the distance between the integers $z$ and $z'$, in this experiment we use a Gaussian kernel $k(z, z') = \exp(-(z-z')^2/(2\sigma^2))$ and set the bandwidth, $\sigma$, using the median heuristic \cite{Gretton2012} applied to the data $y$.
\par
The training performance of proximal gradient ascent and SGLD is again plotted on the bottom row of Figure~\ref{fig:conv_plots}. We observe that the deep circuit fits to the data very well and quickly even with 45\% of the gates removed. The shallow circuit takes longer to fit although this is mitigated by the noise in SGLD - perhaps helping the parameters to more quickly escape a difficult region in the initialisation around zero.

In Figure~\ref{fig:stamps_sim_data_plots}, we generate a simulated dataset of $10^3$ samples using converged parameters and visually compare with the true data. We observe that the Laplace prior approach has fitted the data well with up to 30\% of the gates removed in the shallow circuit, and up to 45\% for the deep circuit. For the SGLD parameters, we take a different approach where we simulate our dataset by taking 100 samples from each parameter along the training trajectory (after discarding a burn-in of 400 samples), so-called ergodic averages. We observe that this approach provides an implicit regularisation and produces consistent simulated data even in the large noise setting $\beta = 10^3$.
\par

\begin{figure}
    \centering
    \includegraphics[width=\linewidth]{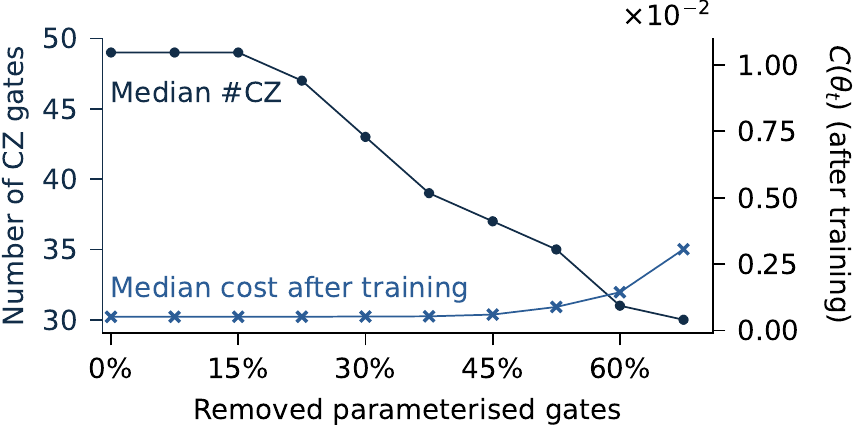}
    \caption{Number of $CZ$ gates for the circuit in \figureautorefname~\ref{fig:circ}, $L=7$, when compiled after proximal gradient ascent training for varying regularisation strength in the generative modelling experiment (8 qubits). Final training cost presented in blue and on right axis. Results are after 1000 iterations of proximal gradient ascent and are repeated across 20 random initialisations with median displayed.}
    \label{fig:zgate_plots}
\end{figure}

In Figure~\ref{fig:zgate_plots}, we depict the number of $CZ$ gates that are cancelled out during compilation of the circuit in Figure~\ref{fig:circ} due to the regularisation of the Laplace prior and subsequent removal of $R_x$ and $R_z$ gates. We see that a significant proportion of $CZ$ gates are removed when more than 40\% of the number of the rotational parameters are removed although when more than 50\% are removed we start to take a hit and suffer poorer performance on the trained cost.
\par

\begin{figure}
    \centering
    \includegraphics[width=\linewidth]{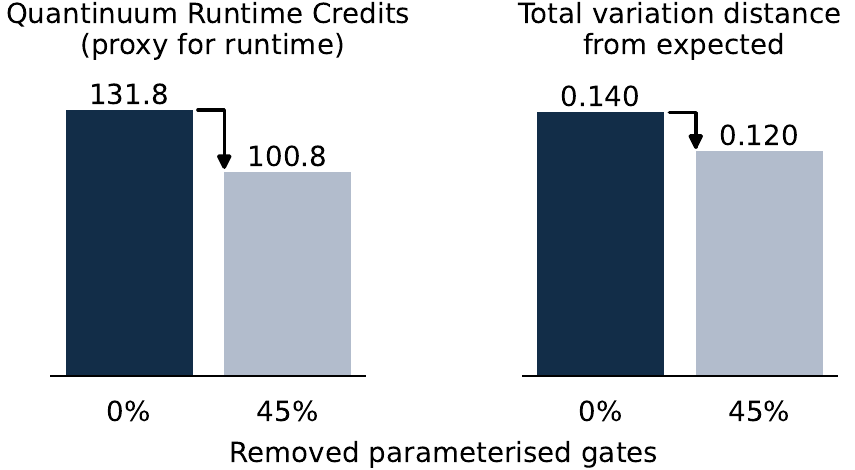}
    \caption{Runtime credits (a proxy for execution time) and total variation distance for experiments on the Quantinuum H1-2 quantum computer (1000 shots) for 0\% and 45\% of parameterised gates automatically removed by the proximal gradient ascent algorithm. Total variation distance is measured from the expected sampling distribution with no hardware noise. Circuits are compiled for and run on H1-2 after training on a simulator. These two circuits provide equivalent MMD performance as seen in Figures \ref{fig:conv_plots}, \ref{fig:stamps_sim_data_plots} and \ref{fig:zgate_plots}. Results are averaged over 5 independent hardware runs.}
    \label{fig:quant_sims}
\end{figure}

Finally, in Figure~\ref{fig:quant_sims}, we take two instances of the trained circuits with depth $L=7$ and compare the sampling cost and hardware noise on the Quantinuum H1-2 trapped-ion quantum computer~\cite{pinoDemonstrationTrappedionQuantum2021} after compiling via tket \cite{Sivarajah2020}. Observe that the circuit with 45\% of paramterised gates removed is significantly cheaper and less noisy than its full parameter equivalent. We also remark that these benefits are multiplied by a factor of $2K$ in each gradient calculation if we employ the parameter shift rule.

\section{Discussion}\label{sec:disc}

In this article we have described a very general Bayesian framework for the probabilistic treatment of variational quantum circuits. By considering the probability distribution $\pi(\theta) \propto p(\theta)\exp(-\beta C(\theta))$ we generalise the existing optimisation framework as a special case where a uniform prior $p(\theta) \propto 1$ is used implicitly and gradient ascent techniques are applied to find a maximum a posteriori estimate $\theta_\text{MAP} = \text{argmax}_\theta \pi(\theta)$. We move beyond the uniform prior and show how a Laplace distribution can be used to enforce customisable levels of sparsity in the parameter $\theta_\text{MAP}$. This dimension reduction has benefits including faster sampling and reduced hardware noise, as well as potential trainability benefits for large problems. We additionally detail how to generate a Monte Carlo approximation that is asymptotically unbiased for the posterior $\pi(\theta)$ via an application of stochastic gradient Langevin dynamics. A characterisation of the posterior beyond a point estimate is highly desirable for landscapes exhibiting complex contours and local optima, as demonstrated by the weighted max-cut and transverse field Ising model experiments. 
\par
This Bayesian perspective leaves many questions for future research, of which we will describe a few here.
\par
Both of the described inference algorithms bring with them an additional tuning parameter, the regularisation strength $\alpha$ in the case of proximal gradient ascent and the level of noise added $\beta$ in the case of SGLD. We described an adaptive method for the regularisation strength $\alpha$, however this approach simply helps via the intuitive nature of deciding a priori how many parameters to remove, leaving an alternative tuning parameter. Another approach would be to put a prior (such as a gamma or log-normal distribution) on $\alpha$ or $\beta$ and include them in the inference procedure.
\par
An extremely successful modification of gradient descent in the case of stochastic gradients is the addition of momenta \cite{Nesterov1983, Kingma2015}. Indeed, a compelling future direction is the extension to proximal gradient ascent with momenta \cite{Qunwei2017} and the sampling analogue of momenta, underdamped Langevin dynamics \cite{Chen2014, Leimkuhler2016}. Whilst there is also the opportunity to incorporate second-order information into the Bayesian inference regimes via the use of a preconditioner \cite{Ma2015, Parikh2014}.
\par
Additionally, we have only investigated point estimate and Monte Carlo approximations to the posterior. A natural next step would be to consider a variational inference approach \cite{Blei2017}, although care would need to be taken when constructing a variational family of distributions that are well-defined for angular parameters.
\par
The posterior $\pi(\theta)$ represents an instance of generalised Bayesian inference \cite{Bissiri2016, Knoblauch2019}, there are indeed alternative posterior formulations providing a probabilistic interpretation of uncertainty over $\theta$. A particularly compelling alternative approach corresponds to \textit{approximate Bayesian computation} (e.g. \cite{Beaumont2019}) which has desirable asymptotics and permits a Metropolis-Hastings accept-reject step. This approach is expanded on in \appendixautorefname~\ref{abc} and represents a significant reformulation of the cost function and inference procedure although remains an intriguing future direction nonetheless.
\par
One of the major concerns for the utility of variational quantum algorithms is that of trainability in large circuits and the so-called barren plateau phenomenon \cite{mccleanBarrenPlateausQuantum2018} where the gradients of randomly initialised circuits vanish exponentially as the number of qubits increases. It is a natural question to consider whether the choice of prior can mitigate the barren plateau phenomenon either via automatic dimension reduction of the Laplace prior or by introducing correlations amongst parameters \cite{volkoffLargeGradientsCorrelation2021} via e.g. a (correlated) von Mises prior distribution \cite{Mardia2000}.
\par
A significant motivation for regularisation in classical statistics and machine learning is that of generalisation. $\ell_1$ regularisation (without the proximal approach) is investigated for quantum supervised learning with preliminary mixed results in \cite{Qian2021}, it would be intriguing to investigate whether alternative priors could help quantum circuits avoid overfitting. In the same vein, classical Bayesian deep learning \cite{Wilson2020} (where the posterior samples are preferred over a point estimate) represents a compelling approach to improving generalisation, strongly motivating the extension of the quantum Bayesian learning framework described here to supervised learning.

\begin{acknowledgements}
We thank Kirill Plekhanov for fruitful discussions, feedback on the manuscript and help with the implementation. We thank Stephen Ragole and Richie Yeung for providing feedback on the manuscript.
\end{acknowledgements}

\bibliography{main}

\appendix
\section{A Note on Approximate Bayesian Computation}\label{abc}

A suitably decaying stepsize schedule is one approach to correct for the discretisation error in the Langevin proposal \eqref{langevin_em}. Another possibility is to keep the stepsize constant and apply a Metropolis-Hastings accept-reject step, where a sample from a proposal distribution $q(\theta' \mid \theta_{t-1})$ is accepted $\theta_t = \theta'$ with probability
\begin{equation*}
    \alpha_t = \min(1, r_t), \qquad \text{where} \qquad r_t = \frac{\pi(\theta') q(\theta_{t-1} \mid \theta')}{\pi(\theta_{t-1}) q(\theta' \mid \theta_{t-1})},
\end{equation*}
otherwise the previous is duplicated $\theta_t = \theta_{t-1}$.
\par
The $\pi(\theta)$ evaluations within $r_t$ only need to be up to normalising constant and can even be replaced with an unbiased estimate \cite{Andrieu2009}. However, in the present formulation \eqref{posterior} we only have access to unbiased estimates of the (unnormalised) log density $\log p(\theta) - \beta C(\theta)$, which cannot be easily translated into an unbiased estimate of the required $p(\theta)\exp(-\beta C(\theta))$.
\par
An alternative formulation instead works directly in the density space. Denote the quantum circuit as a conditional distribution $p(z \mid \theta)$ and a weighting function $k(z)$ that is large when the output $z$ is accurate/desirable and small when $z$ is inaccurate/undesirable. Note that we cannot evaluate $p(z \mid \theta)$ but can extract unbiased estimates for quantities of the form $\mathbb{E}_{p(z \mid \theta)}[f(z)]$ and $\nabla_\theta \mathbb{E}_{p(z \mid \theta)}[f(z)]$. This formulation falls within the field of \textit{approximate Bayesian computation} (ABC) \cite{Beaumont2019} where we target the extended distribution
\begin{equation*}
    \pi_\text{ABC}(\theta, z) \propto p(\theta) p(z \mid \theta) k(z).
\end{equation*}
Discarding the simulated output $z$ amounts to marginalisation
\begin{align*}
    \pi_\text{ABC}(\theta) &\propto \int p(\theta) p(z \mid \theta) k(z) dz,
    \\
    &\propto p(\theta) \mathbb{E}_{p(z \mid \theta)}[k(z)].
\end{align*}
The extended distribution permits a Metropolis-Hastings step \cite{Andrieu2009} and a Langevin proposal, although the Langevin proposal requires the gradient
\begin{align*}
    \nabla_\theta \log \pi_\text{ABC}(\theta) &= \nabla_\theta \log p(\theta) + \nabla_\theta \log \mathbb{E}_{p(z \mid \theta)}[k(z)],
    \\
    &= \nabla_\theta \log p(\theta) + \frac{\nabla_\theta \mathbb{E}_{p(z \mid \theta)}[k(z)]}{\mathbb{E}_{p(z \mid \theta)}[k(z)]},
\end{align*}
which differs from \eqref{log_grad_post}.
\par
In statistics or machine learning settings, e.g. Section~\ref{gen_mod}, we have $k(z) = k(z, y)$ encouraging the output $z$ to be similar to a given dataset $y$. Here $\pi_\text{ABC}(\theta)$ has the desirable property that as $k(z, y) \to \delta(z \mid y)$ we get $\pi_\text{ABC}(\theta) \to p(\theta \mid z) \propto p(\theta) p(z \mid \theta)$ which is in some sense the \textit{ideal} posterior. ABC targets an extended distribution and this has largely limited the approach to low dimensional settings, however this could be mitigated by the use of the gradients above (ABC is usually gradient-free) or by accepting a bias \cite{Duffield2022}. However, this alternative ABC formulation loses the seamless transition from existing optimisation-based variational quantum algorithms, Algorithm~\ref{ga}; a numerical investigation into $\pi_\text{ABC}(\theta)$ is therefore left for future work.

\end{document}